\documentclass[aps,prl,twocolumn,superscriptaddress,citeautoscript]{revtex4-2}

\pdfoutput=1
\usepackage{graphicx}
\usepackage{amsmath, amsfonts}
\usepackage{amssymb}
\usepackage[colorlinks=true,linkcolor=blue,citecolor=blue]{hyperref}
\usepackage[utf8]{inputenc}

\begin{document}

\title{Picosecond Time-Scale Resistive Switching Monitored in Real-Time}

\author{Mikl\'{o}s Csontos}
\affiliation{Institute of Electromagnetic Fields, ETH Zurich, Gloriastrasse 35, 8092 Zurich, Switzerland}

\author{Yannik Horst}
\affiliation{Institute of Electromagnetic Fields, ETH Zurich, Gloriastrasse 35, 8092 Zurich, Switzerland}

\author{Nadia Jimenez Olalla}
\affiliation{Institute of Electromagnetic Fields, ETH Zurich, Gloriastrasse 35, 8092 Zurich, Switzerland}

\author{Ueli Koch}
\affiliation{Institute of Electromagnetic Fields, ETH Zurich, Gloriastrasse 35, 8092 Zurich, Switzerland}

\author{Ivan Shorubalko}
\affiliation{Transport at Nanoscale Interfaces Laboratory, Empa, Swiss Federal Laboratories for Materials Science and Technology, 8600 D\"ubendorf, Switzerland}

\author{Andr\'{a}s Halbritter}
\affiliation{Department of Physics, Institute of Physics, Budapest University of Technology and Economics, M\H{u}egyetem rkp. 3, H-1111 Budapest, Hungary}
\affiliation{ELKH-BME Condensed Matter Research Group, M\H{u}egyetem rkp. 3, H-1111 Budapest, Hungary}

\author{Juerg Leuthold}
\affiliation{Institute of Electromagnetic Fields, ETH Zurich, Gloriastrasse 35, 8092 Zurich, Switzerland}

\begin{abstract}
The resistance state of filamentary memristors can be tuned by relocating only a few atoms at interatomic distances in the active region of a conducting filament. Thereby the technology holds promise not only in its ultimate downscaling potential and energy efficiency but also in unprecedented speed. Yet, the breakthrough in high-frequency applications still requires the clarification of the dominant mechanisms and inherent limitations of ultra-fast resistive switching. Here we investigate bipolar, multilevel resistive switchings in tantalum pentoxide based memristors with picosecond time resolution. We experimentally demonstrate cyclic resistive switching operation due to 20~ps long voltage pulses of alternating polarity. Through the analysis of the real-time response of the memristor we find that the set switching can take place at the picosecond time-scale where it is only compromised by the bandwidth limitations of the experimental setup. In contrast, the completion of the reset transitions significantly exceeds the duration of the ultra-short voltage bias, demonstrating the dominant role of thermal diffusion and underlining the importance of dedicated thermal engineering for future high-frequency memristor circuit applications.
\end{abstract}

\date{\today}
\maketitle

\textbf{1. Introduction}
\vspace{5mm}

Due to the self-assembling, atomic-scale formation or destruction of their conducting channels, filamentary memristors \cite{Strachan2010,Miao2011,Lee2011a,Liu2012a,Yang2012,Park2013,Yang2014,Cheng2019} simultaneously exhibit the key properties seeked for the hardware elements of future computational architectures \cite{Burr2017,Zidan2018a,Ielmini2018,Xia2019,Sebastian2020}. Ultimate downscalability is granted by the single atom level control of the filament formation \cite{Schirm2013,Torok2020}. In addition to sub-nanometer filament diameters, devices with 2$\times$2~nm$^{2}$ electrode cross-section areas were also demonstrated \cite{Pi2019}. The analog tunability of the filaments were exploited in large crossbar arrays performing vector-matrix multiplications in only one computational step at greatly reduced power consumption \cite{Bayat2018,Zidan2018b,Li2018a,Lin2020}. Furthermore, the bias voltage dependent, highly nonlinear dynamical properties of the resistive switching \cite{Waser2009,Gubicza2015a} were employed to realize reservoir computing architectures based on only a few physical nodes \cite{Moon2019} as well as to prevent undesired leakage currents in memristor crossbar arrays \cite{Rao2022}.

It is also conceivable, that the relocation of only a few atoms upon resistive switching must be feasible at an extremely fast pace, qualifying memristors to enter the frequency domain of cutting edge telecommunication technologies. Oxide and nitride based devices have indeed been tested to fast resistive switching induced by individual voltage pulses \cite{Torrezan2011,Strachan2011,Pickett2012,Choi2016a,Marchewka2016,Molnar2018,Bottger2020} as short as 50~ps \cite{Witzleben2020,Witzleben2021a}, while circuit design considerations \cite{Strachan2013,Bottger2020,Witzleben2021b} and molecular dynamical simulations \cite{Emboras2018} predict a possibility for up to two orders of magnitude faster operation, bringing even Terahertz frequency applications in sight. Yet, to date memristors have merely been implemented as static switches in broad-band circuits \cite{Pi2015}, where their potential of ultra-fast switching operation was not exploited. In order to optimize memristor devices for high-speed operation in real circuits, the identification of the dominant switching mechanisms as well as the microscopic and design-related speed limitations is inevitable. Beyond verifying the equilibrium resistance states between ultra-fast programming voltage pulses, this requires the real-time analysis of the switching dynamics also during the pulses.

\begin{figure*}[t!]
     \includegraphics[width=2\columnwidth]{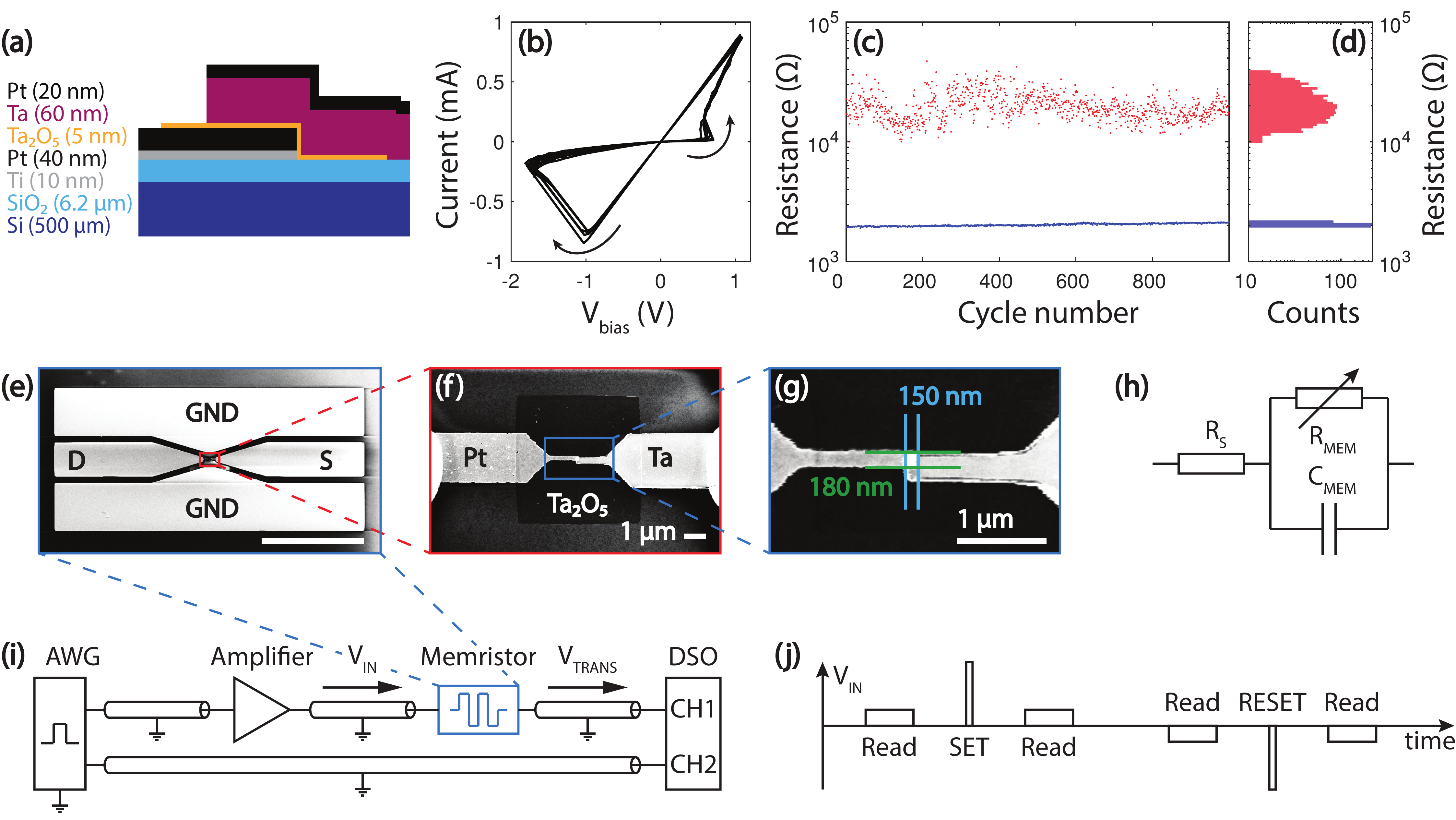}
     \caption{Device structure and the fast switching setup. (a) The (not to scale) vertical cross-section of the memristor device. The layer sequence and the thicknesses are indicated on the left. (b) 10 consecutive $I(V)$ traces acquired with an $f$ = 1~Hz frequency and $V_{\rm drive}^{0}$ = 2~V amplitude triangular voltage signal. The arrows indicate the direction of the hysteresis. The dc setup is explained in the Methods section. (c) Illustration of the stability of the resistive switching operation over 1000 dc $I(V)$ cycles recorded at identical conditions to those in (b). The $R_{\rm LRS}$ (low resistance state, blue) and $R_{\rm HRS}$ (high resistance state, red) resistance values were evaluated from the zero bias slopes of the hysteretic $I(V)$ traces. (d) Histograms of the $R_{\rm LRS}$ (blue) and $R_{\rm HRS}$ (red) data shown in (c). (e) SEM image of the device embedded in a co-planar waveguide structure. The white scale bar in the lower right indicates 100~$\mu$m. The width of the planar gap is 4~$\mu$m. (f) A magnified view of the memristor device with the Pt bottom electrode on the left and the Ta top electrode on the right. (g) A further zoom into the device area shows the measured overlap area of 180~nm $\times$ 150~nm between the top and bottom electrodes. (h) The equivalent circuit of the memristor device accounting for the lead resistance of the device as well as the contact resistance of the needle probes ($R_{\rm S}$), the tunable device resistance ($R_{\rm MEM}$) and the device capacitance arising from the vertical metal-insulator-metal stack ($C_{\rm MEM}$). (i) Schematics of the high-speed setup. The memristor device is exposed to the $V_{\rm IN}$ amplified voltage pulses provided by the arbitrary waveform generator (AWG). The $V_{\rm TRANS}$ transmitted voltage pulses are measured by a digital storage oscilloscope (DSO). The second output channel of the AWG is used to trigger the DSO. (j) The (not to scale) schematics of the voltage pulse sequence repeated during the high-speed switching experiments.}
     \label{setup.fig}
\end{figure*}

Here we present alternating resistive switching and multilevel programming in tantalum-pentoxide based memristors due to 20~ps full width at half maximum (FWHM) set and reset pulses, confirmed by the evaluation of the steady states before and after the programming pulses. More importantly, we move beyond the common approach of verifying the effect of the programming pulses in the established steady state only: We have developed an optimized sample design which enabled us to monitor the dynamics of the resistive switching also during the ultra-short programming voltage pulses. Thereby set times well below 20~ps pulse durations could be measured at picosecond resolution. In addition, a thermal delay of the reset transition, which exceeds the reset voltage pulse duration, was discovered. The latter also allows the experimental separation of electric field induced and diffusion driven resistive switching mechanisms. The intrinsic, structural and instrumental speed limitations of resistive switching are evaluated both in the set and reset direction, facilitating low-power memristor circuits for telecommunication technologies as well as ultra-fast neuromorphic computing applications.

\vspace{5mm}
\textbf{2. Results and Discussion}
\vspace{5mm}

Our report is structured as follows. First, the memristor structure and the principles of the transmission measurements of fast voltage pulses are summarized. Next, following the more traditional approach, alternating resistive switchings due to ultra-fast voltage pulses down to 20~ps FWHM are evaluated in the steady high resistance (HRS) and low resistance (LRS) states, exhibiting multilevel programming and satisfying the `voltage-time dilemma' \cite{Waser2009,Gubicza2015a}. Moving an important step further, we assess the criteria of experimentally resolving resistive switchings \emph{during} the ultra-fast voltage pulses and, accordingly, demonstrate set switching times at the picosecond time-scale. Finally, also by analyzing the time dependence of the transmitted voltage during the reset pulses, we provide experimental evidence that the thermally driven reset transition can take place at longer time-scales than the actual reset voltage pulse width.

\vspace{5mm}
\textbf{2.1 Memristor Structure and DC Characterization}
\vspace{5mm}

The device structure and the individual thicknesses of the vertically stacked layers are shown in Fig.~\ref{setup.fig}(a). The Pt bottom electrode was evaporated onto a standard Si/SiO$_{2}$ wafer using a thin Ti adhesive layer. The 5~nm thick Ta$_{2}$O$_{5}$ switching layer, the Ta top electrode and the Pt cap layer were deposited by high-power impulse magnetron sputtering. The lateral patterning of the devices was carried out by standard electron beam lithography and lift-off. Figure~\ref{setup.fig}(e) illustrates the coplanar waveguide structure hosting the device under test. For this purpose 200~nm thick Au electrodes were evaporated using a resist mask defined by standard optical lithography. The subsequentially magnified electron microscope images in Figs.~\ref{setup.fig}(f) and (g) provide an insight into the electrode layout and the 180~nm $\times$ 150~nm overlaying area of the top and bottom electrodes. Further fabrication details and considerations on the sample layout are detailed in the Methods section.

The dc performance of the Ta/Ta$_{2}$O$_{5}$/Pt devices are summarized in Fig.~\ref{setup.fig}(b-d), exemplifying the typical, hysteretic current-voltage [$I(V)$] traces, their excellent cycle-to-cycle reproducibility and routinely obtained endurance over 1000 consecutive operation cycles. The statistical analysis of the zero bias resistance values in Fig.~\ref{setup.fig}(d) reveals a narrow distribution of the low resistance states (LRS) around $R_{\rm LRS}\approx$2~k$\Omega$ and a broader set of the high resistance states (HRS) with $R_{\rm HRS}$ in the 10\,--\,30~k$\Omega$ range. Note that this resistance regime, where the conducting filaments are only partially destructed also in the HRS while the LRS traces show good linearity up to the switching threshold, is not only the favorable choice for analog multilevel operations \cite{Li2018a} but is also ideal for the optimum resolution in the transmission experiments of high-speed voltage pulses outlined in the following.

\vspace{5mm}
\textbf{2.2 Resistive Switching Due to Ultra-Short Voltage Pulses}
\vspace{5mm}

The schematics of the fast pulsing setup is shown in Fig.~\ref{setup.fig}(i). A programmable 100~GSa/s DAC board system was utilized as an arbitrary waveform generator (AWG) capable of firing voltage pulses down to 20~ps FWHM at 20~ps rise time. The voltage output of the AWG was further amplified up to $\pm$4~V peak values by a broadband amplifier module specified to a 65~GHz analog bandwidth. The voltage signals propagated in 30~cm long, 65~GHz bandwidth, 50~$\Omega$ terminated coaxial RF cables. The memristor sample was engaged by two 67~GHz bandwidth triple probes in a custom-built probe station. The $V_{\rm TRANS}$ transmitted voltage signals were recorded by a 50~$\Omega$ terminated digital storage oscilloscope (DSO) operated at a 256~GSa/s sampling rate and 113~GHz analog bandwidth. A further enhanced, picosecond time resolution was achieved by taking advantage of the high reproducibility of the repeated switching cycles, as discussed later in Section~2.4. The $V_{\rm IN}$ signal was recorded separately by bypassing the memristor sample, the probes and the last cable section.

The equivalent circuit model of the memristor device is drawn in Fig.~\ref{setup.fig}(h). This is composed of a series resistor $R_{\rm S}$, a parallel capacitor $C_{\rm MEM}$ and a resistor $R_{\rm MEM}$ representing the contributions of the lead and probe contact resistances, the parallel capacitance arising from the metal-insulator-metal stack of the memristor device and the memristive resistance, respectively. A numerical analysis revealing the influence of these circuit parameters on the measured $V_{\rm TRANS}$ voltage signal served as a guide to the optimization of the sample layout, as discussed in Section~2.4 and in the Methods section. In short, the real-time monitoring of the memristor's resistive response during the ultra-fast $V_{\rm IN}$ pulses critically relies on the careful minimization of $C_{\rm MEM}$ as well as further environmental parasitic capacitances but, at the same time, also requires the minimization of $R_{\rm S}$.

Finally, the applied pulse scheme is illustrated in Fig.~\ref{setup.fig}(j). In this work we demonstrate subsequent resistive switchings due to ultra-fast set and reset programming pulses without interrupting the measurement by dc cycling of the device. The low-bias read-out of the equilibrium resistive state between the programming pulses was carried out by non-invasive, low-amplitude voltage pulses of 1~ns duration, preceding and following the 20\,--\,500~ps FWHM programming pulses by 10~ns. As a voltage polarity convention, positive bias corresponds to a higher potential applied on the Ta top electrode with respect to the Pt bottom electrode. The period of the set/reset cycle was 5.24~$\mu$s, the programming pulses of alternating sign followed at every 2.62~$\mu$s. In a `single shot' experiment 200\,--\,300 subsequent cycles were executed and recorded at maximum sampling rates.

\begin{figure*}[t!]
     \includegraphics[width=2\columnwidth]{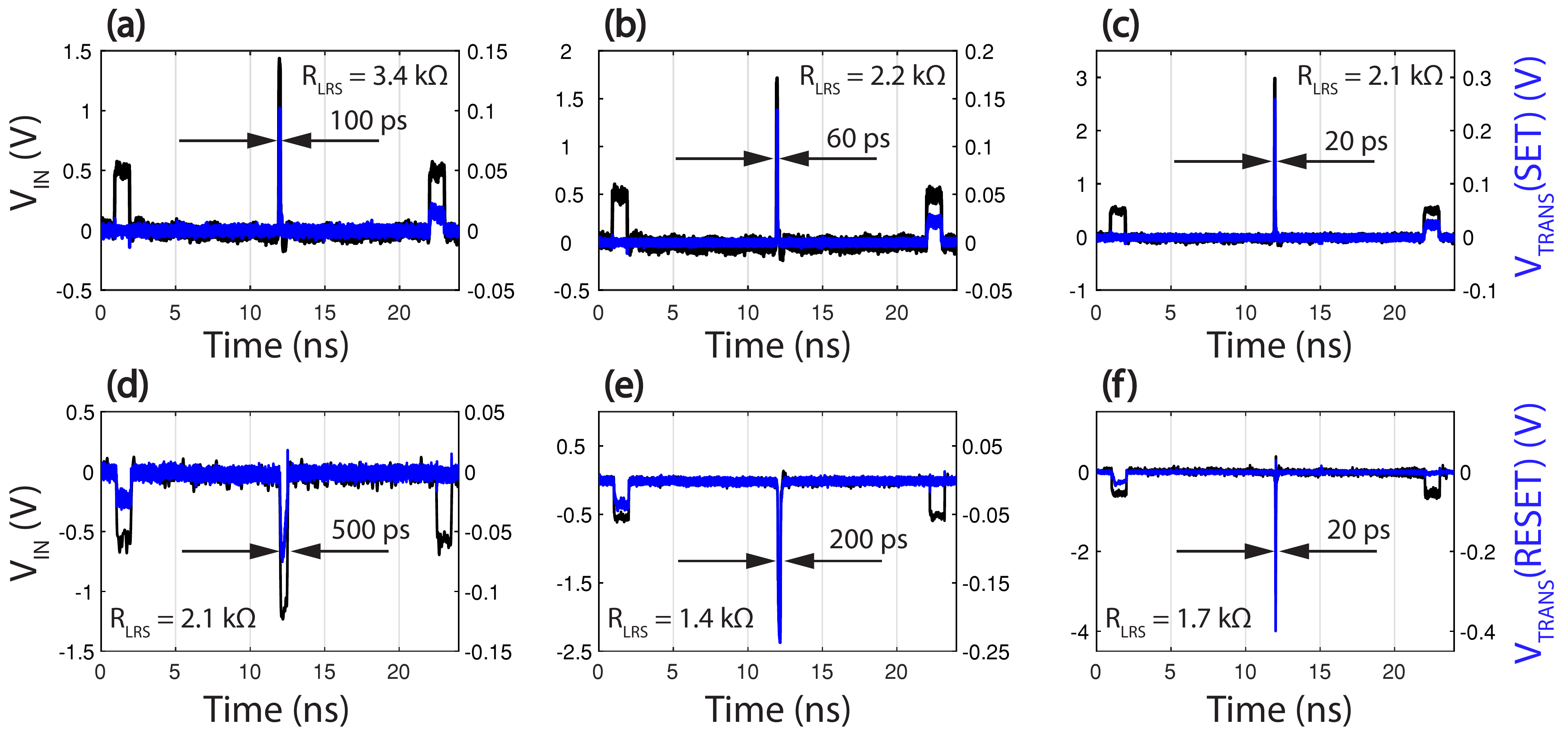}
     \caption{The effect of the fast set and reset pulses are monitored by 1~ns long read pulses in the steady state at various programming pulse durations, confirming resistive switching. (a-c) Set operation due to 100~ps, 60~ps and 20~ps voltage pulses, respectively. (d-f) Reset operation due to 500~ps, 200~ps and 20~ps voltage pulses, respectively. The left axes correspond to $V_{\rm IN}$ (black) whereas $V_{\rm TRANS}(\rm SET)$ (upper panels, blue) and $V_{\rm TRANS}(\rm RESET)$ (lower panels, blue) are scaled on the right axes. The $R_{\rm LRS}$ mean values evaluated by the 1~ns read pulses as well as the FWHM of the programming pulses are labeled on all panels. $R_{\rm HRS} >$ 20~k$\Omega$.}
     \label{ns.fig}
\end{figure*}

The analysis of the device impedance based on the $V_{\rm TRANS}$ signal relies on the solution of the telegraph equations derived for our arrangement \cite{Santa2020} which results in the formula of
\begin{equation}
\frac{V_{\rm TRANS}}{V_{\rm IN}}=\frac{2Z_{0}}{Z_{\rm MEM}+2Z_{0}} \mbox{ ,}
\label{trans.eq}
\end{equation}
where $Z_{\rm MEM}$ is the complex impedance of the memristor represented in Fig.~\ref{setup.fig}(h) and $Z_{0}$=50~$\Omega$ is the wave impedance of the transmission lines. The time delay between the incoming and transmitted pulses arising from the propagation along finite cable lengths is compensated, as detailed in Section~2.4. We note that Eq.~\ref{trans.eq} directly applies to plane waves and wave packages in the presence of a frequency independent, resistive $Z_{\rm MEM}$. When a complex, frequency dependent $Z_{\rm MEM}(f)$ is involved, the calculation of the time dependent $V_{\rm TRANS}$ response requires a Fourier analysis based on Eq.~\ref{trans.eq} and $Z_{\rm MEM}(f)$. Such numerical simulations are presented in the Methods section.

As will be discussed later, the capacitive / resistive character of $Z_{\rm MEM}$ has a profound influence on the time-dependence of $V_{\rm TRANS}$ during the ultra-fast programming pulses. However, during the 1~ns long read pulses a finite capacitive contribution to $V_{\rm TRANS}$ only arises at the steep, 20~ps long pulse edges. Owing to the low, $C_{\rm MEM}$=3~fF capacitance of our devices, $V_{\rm TRANS}$ between the edges of the 1~ns read pulse is exclusively dominated by the effect of $R_{\rm MEM}$. Thereby, according to Eq.~\ref{trans.eq}, the $V_{\rm TRANS}/V_{\rm IN}$ ratio becomes a real-valued number and the steady state resistance can be directly determined as $R_{\rm MEM}=2Z_{0}(V_{\rm IN}/V_{\rm TRANS}-1)$.

In the following, we utilize this approach to evaluate the steady state device resistances before and after the ultra-fast programming pulses. The resistance resolution of this method is limited by the noise floor of the DSO at the voltage range needed to cover the $V_{\rm TRANS}$ signal also during the high-amplitude programming pulses. Consequently, the applied transmission method is able to resolve resistances in the $R_{\rm MEM}\lesssim$20~k$\Omega$ regime with increasing resolution toward lower $R_{\rm MEM}$ values.

The electroforming step preceding the fast pulsing operation did not require the application of any dedicated, current limited dc voltage sweep. Instead, the pristine devices were exposed to the $V_{\rm IN}$ pulsing sequences illustrated in Fig.~\ref{setup.fig}(j) where the amplitude of the set pulse was gradually increased until stable resistive switching was established. After successful electroforming, the amplitude of the set pulse could be reduced again while cyclic resistive switching operation was still maintained until the threshold amplitude corresponding to the actual set pulse width was reached. As a matter of experience, this more gentle, pulse-based electroforming approach resulted in higher endurance and larger $R_{\rm HRS}/R_{\rm LRS}$ ratios.

Set switchings due to programming pulses with 100~ps, 60~ps and 20~ps FWHM and varying amplitudes are demonstrated in Fig.~\ref{ns.fig}(a-c), respectively. Beside the set pulses, the 24~ns time interval of the panels also exhibits the 1~ns FWHM, $V_{\rm IN}$=0.5~V read pulses. The latter two parameters were adjusted in a way that, in the absence of a programming pulse, the read pulses do not induce any detectable resistive switching over 100's of cycles. The time axis of $V_{\rm TRANS}(\rm SET)$ (blue, right axes) is compensated for cable length differences with respect to $V_{\rm IN}$ (black, left axes). The flat response to the first read pulse falls below the noise floor of the measurement confirming the HRS with $R_{\rm HRS}>$20~k$\Omega$ at each displayed case. In contrast, the finite $V_{\rm TRANS}(\rm SET)$ during the second read pulse testifies to the LRS with around 2\,--\,4~k$\Omega$ resistance, as labeled in the panels. Moreover, the finite resistive response during the set pulse indicates that resistive switching took place already within the duration of the latter (not resolved here).

Successful reset switchings triggered by 500~ps, 200~ps and 20~ps long programming pulses were tested in a similar fashion to the set operation utilizing the second half of the scheme in Fig.~\ref{setup.fig}(j), as shown in Fig.~\ref{ns.fig}(d-f), respectively. After verifying that in the absence of a reset pulse they do not induce any reset transitions, 1~ns long and 0.5~V amplitude voltage pulses of negative polarity were utilized for the read operation. The finite $V_{\rm TRANS}(\rm RESET)$ (blue, right axes) values during the first read pulses (black, left axes) clearly demonstrate that the device is prepared in the LRS of $R_{\rm LRS}\approx$2~k$\Omega$, as labeled in the panels. During the shorter programming pulses a predominantly resistive response, characteristic to the LRS is observed (not resolved here), making the more involved time-dependent analysis outlined in the following sections necessary to determine the actual reset times. Nevertheless, the absence of a finite resistive response to the second read pulses unambiguously confirm the transition to the HRS with $R_{\rm HRS}>$20~k$\Omega$.

\vspace{5mm}
\textbf{2.3 Multilevel Programming}
\vspace{5mm}

\begin{figure}[t!]
     \includegraphics[width=\columnwidth]{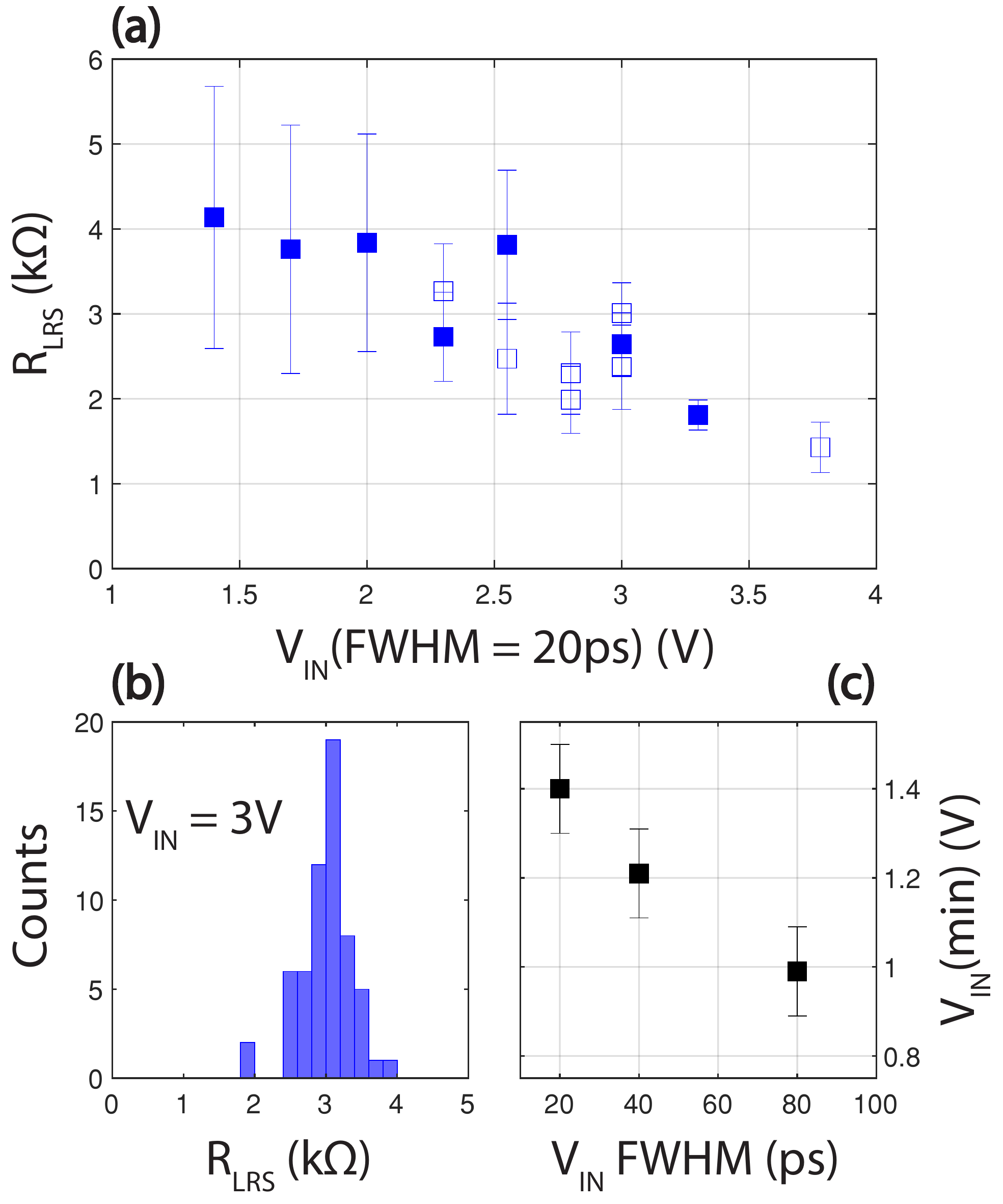}
     \caption{Multilevel programming in the LRS due to 20~ps FWHM set pulses and voltage-time dilemma in the sub-100 picosecond time domain. (a) The tunability of the LRS by 20~ps FWHM set voltage pulses. The filled and empty squares represent $R_{\rm LRS}$ data acquired on two nominally identical devices. The symbols and the error bars correspond to the mean value and standard deviation of the resistance. They were determined by the 1~ns long, 0.5~V amplitude read pulses applied after each set pulse during a sequence of 50-100 alternating set/reset cycles. $R_{\rm HRS} >$ 20~k$\Omega$. (b) A representative $R_{\rm LRS}$ distribution corresponding to $V_{\rm IN}$ = 3~V. (c) The pulse width dependence of the $V_{\rm IN}(\rm min)$ lowest set voltage amplitude which induced resistive switching.}
     \label{multilevel.fig}
\end{figure}

The effect of the 20~ps FWHM set pulse amplitude on the resistive switching is shown in Fig.~\ref{multilevel.fig}. First, $R_{\rm LRS}$ was statistically evaluated based on the resistive, real-valued $V_{\rm TRANS}/V_{\rm IN}$ ratio harvested from the data acquired during the second read pulses over 50-100 consecutive switching cycles. Figure~\ref{multilevel.fig}(a) shows the mean values (square symbols) and the standard deviations (error bars) of the calculated $R_{\rm LRS}$ distributions at each amplitude setting of the 20~ps FWHM set pulses. A representative $R_{\rm LRS}$ histogram corresponding to $V_{\rm IN}$=3~V is illustrated in Fig.~\ref{multilevel.fig}(b). At the minimum $V_{\rm IN}(\rm min)$ amplitude where resistive switching was observed at 20~ps FWHM, the LRS was found to exhibit $R_{\rm LRS}\lesssim$5~k$\Omega$. The further monotonous, up to five-fold decrease in $R_{\rm LRS}$ with increasing $V_{\rm IN}$ shows the potential of multilevel programming using the shortest available set pulses. Figure~\ref{multilevel.fig}(c) summarizes the $V_{\rm IN}(\rm min)$ amplitudes as a function of the set voltage pulse duration in the sub-100~ps time domain. Here the error bars correspond to the discreteness of the amplitude adjustments arising from the 6 bit vertical resolution of the AWG. The observed linear increase of 0.2~V in the minimum amplitude at exponentially shortened pulse durations confirms that the general tendency expressed in the voltage-time dilemma sustains also in the so-far largely unexplored 20\,--\,100~ps regime.

\vspace{5mm}
\textbf{2.4 The Real-Time Monitoring Approach}
\vspace{5mm}

In the previous sections resistive switchings due to ultra-short voltage pulses were demonstrated by investigating the equilibrium states before and after the pulses. However, this approach does not provide a real-time insight into the resistive transitions and, thus, only allows crude estimates of the involved time-scales. In order to take the next step, we analyze the time-dependence of the transmission during the $V_{\rm IN}$ pulses. We show that by an optimized device layout the equivalent circuit parameters introduced in Fig.~\ref{setup.fig}(h) can be adjusted such that a high contrast can be achieved between two reference traces, the non-switching HRS and LRS responses denoted in the following by $V_{\rm TRANS}(\rm HRS)$ and $V_{\rm TRANS}(\rm LRS)$, respectively. The resistive switching response will be quantitatively compared to these two references, all measured at identical $V_{\rm IN}$ bias. This comparison allows the deduction of the resistive switching time-scales.

\begin{figure}[t!]
     \includegraphics[width=\columnwidth]{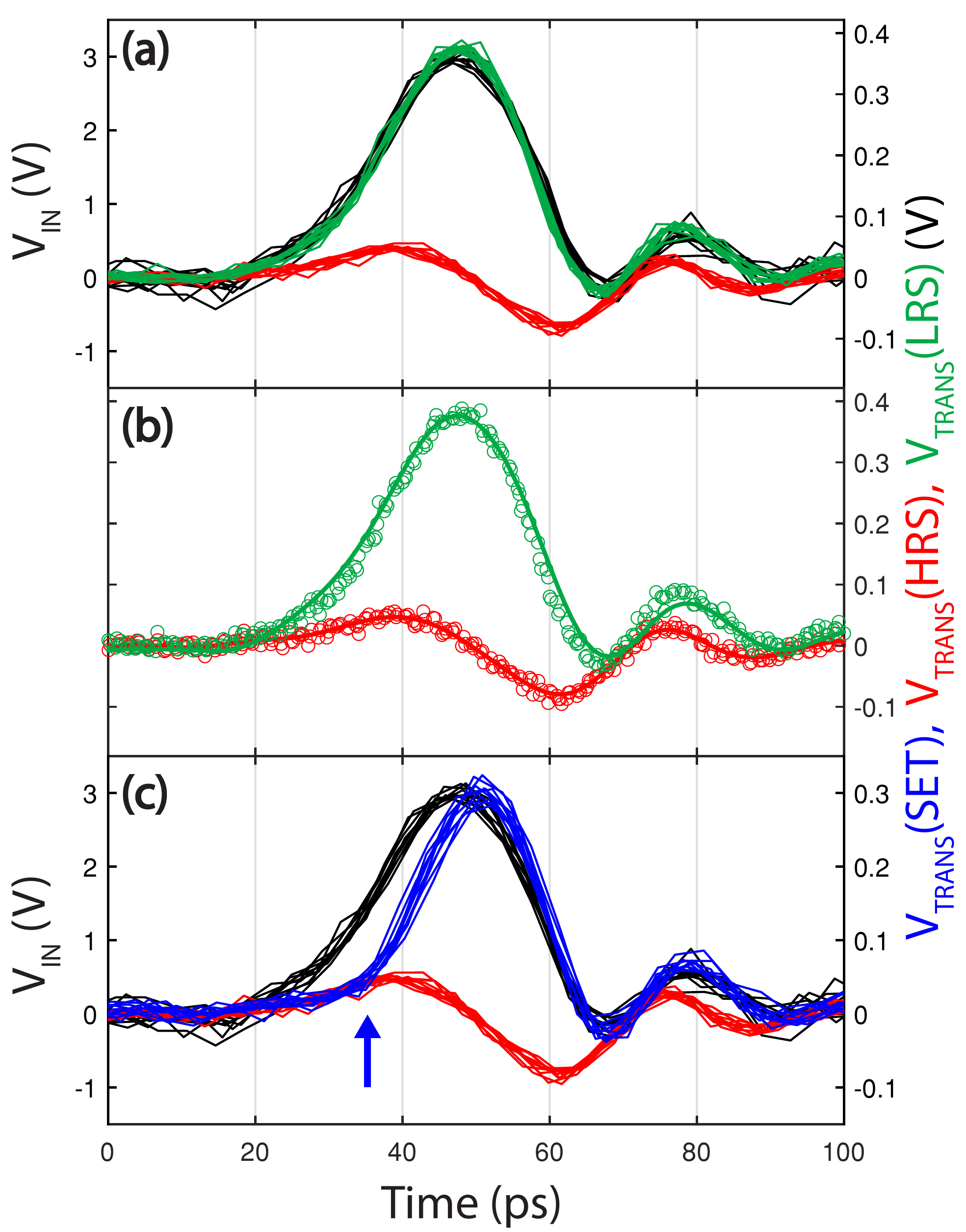}
     \caption{Analysis of the set switching during the programming pulse. In the non-switching LRS, the $V_{\rm TRANS}(\rm LRS)$ response (green, right axes) is proportional to $V_{\rm IN}$ (black, left axes). In the non-switching HRS, the $V_{\rm TRANS}(\rm HRS)$ response (red, right axes) is not proportional to $V_{\rm IN}$ due to the dominant capacitive contribution, but it is sufficiently suppressed due to the careful minimization of $C_{\rm MEM}$. The $V_{\rm TRANS}(\rm SET)$ set switching response (blue, right axis) exhibits a crossover between the above two reference traces which allows the identification of the switching time. (a) The overlaying of 10 consecutive, 20~ps FWHM $V_{\rm IN}$ set voltage pulses (black, left axis) as well as the corresponding, non-switching $V_{\rm TRANS}(\rm HRS)$ (red, right axis) and non-switching $V_{\rm TRANS}(\rm LRS)$ (green, right axis) responses demonstrates the cycle-to-cycle reproducibility and the resulting enhanced time resolution. (b) The empty symbols correspond to the same data as in (a). The solid lines show the simulated transmissions calculated from the measured $V_{\rm IN}$ signal using the fixed $R_{\rm S}$ = 350~$\Omega$, $C_{\rm MEM}$ = 3~fF and $R_{\rm MEM}$ corresponding to HRS (red) and LRS (green), in accordance with the equivalent circuit shown in Fig.~\ref{setup.fig}(h). More details about the transmission simulations are provided in the Methods section. (c) $V_{\rm IN}$ (black, left axis) and $V_{\rm TRANS}(\rm HRS)$ (red, right axis) as in (a). The $V_{\rm TRANS}(\rm SET)$ traces (blue, right axis) show 10 consecutive, overlayed set switching response pulses of the same device, demonstrating the real-time resolved crossover between the corresponding initial HRS and final LRS. The blue arrow marks the starting of the set transition.}
     \label{overlay.fig}
\end{figure}

For the ultra-fast $V_{\rm IN}$ pulses the transmitted signal is determined by the full $Z_{\rm MEM}$ complex impedance of the device according to Eq.~\ref{trans.eq}. A quantitative analysis about the tendencies on $C_{\rm MEM}$, $R_{\rm MEM}$ and $R_{\rm S}$ at 20~ps FWHM $V_{\rm IN}$ pulses is given in the Methods section. In essence, at sufficiently low $R_{\rm MEM}$ and $C_{\rm MEM}$, $V_{\rm TRANS}(\rm LRS)$ is dominated by the resistive response of the device, resulting in a real-valued $V_{\rm TRANS}/V_{\rm IN}$ ratio as illustrated by the black and green traces in Fig.~\ref{overlay.fig}(a). In contrast, at $R_{\rm MEM}>$20~k$\Omega$, $V_{\rm TRANS}(\rm HRS)$ exhibits a merely capacitive character which is highly sensitive to $C_{\rm MEM}$ but much less to $R_{\rm MEM}$ and $R_{\rm S}$, as shown by the red trace in Fig.~\ref{overlay.fig}(a). The duration of the resistive switching transition is identified as the crossover period of the $V_{\rm TRANS}(\rm SET)$ or $V_{\rm TRANS}(\rm RESET)$ switching response between the two reference traces, as demonstrated in Fig.~\ref{overlay.fig}(c).

Figure~\ref{overlay.fig}(a) illustrates 10 overlayed, highly reproducible consecutive cycles of the measured $V_{\rm TRANS}(\rm HRS)$ (red, right axis) and $V_{\rm TRANS}(\rm LRS)$ (green, right axis) responses due to 20~ps FWHM $V_{\rm IN}$ (black, left axis) set pulses. First, $V_{\rm TRANS}(\rm HRS)$ was recorded in the pristine state, at a set pulse amplitude below the critical amplitude where electroforming occurs but above the minimum threshold of resistive switching in the electroformed state. Although the $\gg$M$\Omega$ pristine state is not to be confused with the $>$20~k$\Omega$ HRS, due to the dominating effect of $C_{\rm MEM}$ at higher $R_{\rm MEM}$ values, they result in experimentally identical $V_{\rm TRANS}$ responses. Thus, the transmission recorded in the pristine state provides a proper, non-switching HRS reference for our time domain analysis. After successful electroforming at a higher voltage level, the switching response was acquired at the reduced set pulse amplitude corresponding to the previously recorded HRS response. Finally, the non-switching LRS response was recorded at an identical set pulse amplitude by reducing the reset pulse amplitude, shown in Fig.~\ref{setup.fig}(j), to zero.

The determination of the relative timing between the $V_{\rm TRANS}$ and the corresponding $V_{\rm IN}$ time traces is a crucial step in the deduction of the resistive switching time-scales below the duration of the ultra-fast pulses and, thus, has to be carried out at utmost accuracy. Our procedure is based on consistently (i) matching the rising edges of the read pulse and of the corresponding LRS response and (ii) matching the experimental and simulated LRS and HRS responses to the corresponding $V_{\rm IN}$ signal during the programming pulse. The details of such $V_{\rm TRANS}$ simulations are discussed in the Methods section. 

The high cycle-to-cycle reproducibility of the overlayed 10 consecutive periods of $V_{\rm IN}$, $V_{\rm TRANS}(\rm HRS)$, $V_{\rm TRANS}(\rm LRS)$ and $V_{\rm TRANS}(\rm SET)$ demonstrated in Fig.~\ref{overlay.fig} is a key to the enhanced time resolution of the transmitted voltage response: while the sampling rate of the DSO facilitates the acquisition of one data point at every 3.9~ps, the subsequent pulses are sampled at different relative timings due to the period of the $V_{\rm IN}$ cycle which is a non-integer multiple of the instrumental sampling interval. Thus, by overlaying the subsequent pulses, the effective time resolution can be greatly enhanced until the noise limit of the DSO is reached. The enhanced time resolution lies in the heart of our time dependent analysis of $V_{\rm TRANS}$ as it enables the monitoring of $<$20~ps long resistive switching transitions at the picosecond time-scale. Consequently, throughout the following analy\-sis of the set and reset responses, 10 consecutive cycles were overlayed, averaged and smoothed.

By the optimization of the device layout (see in Methods) a dominantly resistive character of $V_{\rm TRANS}(\rm LRS)$ could be achieved, as evidenced by the perfect overlap between the adequately scaled $V_{\rm IN}$ and $V_{\rm TRANS}(\rm LRS)$ traces in Fig.~\ref{overlay.fig}(a). At the same time, $V_{\rm TRANS}(\rm HRS)$ exhibits an entirely capacitive character corresponding to $C_{\rm MEM}$=3~fF, while the magnitude of $V_{\rm TRANS}(\rm HRS)$ is kept low. Thereby the desired, high contrast between the reference HRS and LRS responses is established. The excellent agreement between the experimental datapoints [same as in Fig.~\ref{overlay.fig}(a)] and the simulated time-dependent transmissions is presented in Fig.~\ref{overlay.fig}(b). Assuming a dielectric constant of $\epsilon_{r}$=23 for Ta$_{2}$O$_{5}$ \cite{Kim2000}, the deduced $C_{\rm MEM}$=3~fF value is a factor of $\approx$2 higher compared to the prediction of the plate capacitor model taking into account the actual geometrical parameters of our device. We attribute this small difference to the stray capacitance of the electrode and probe arrangements.

\begin{figure*}[t!]
     \includegraphics[width=2\columnwidth]{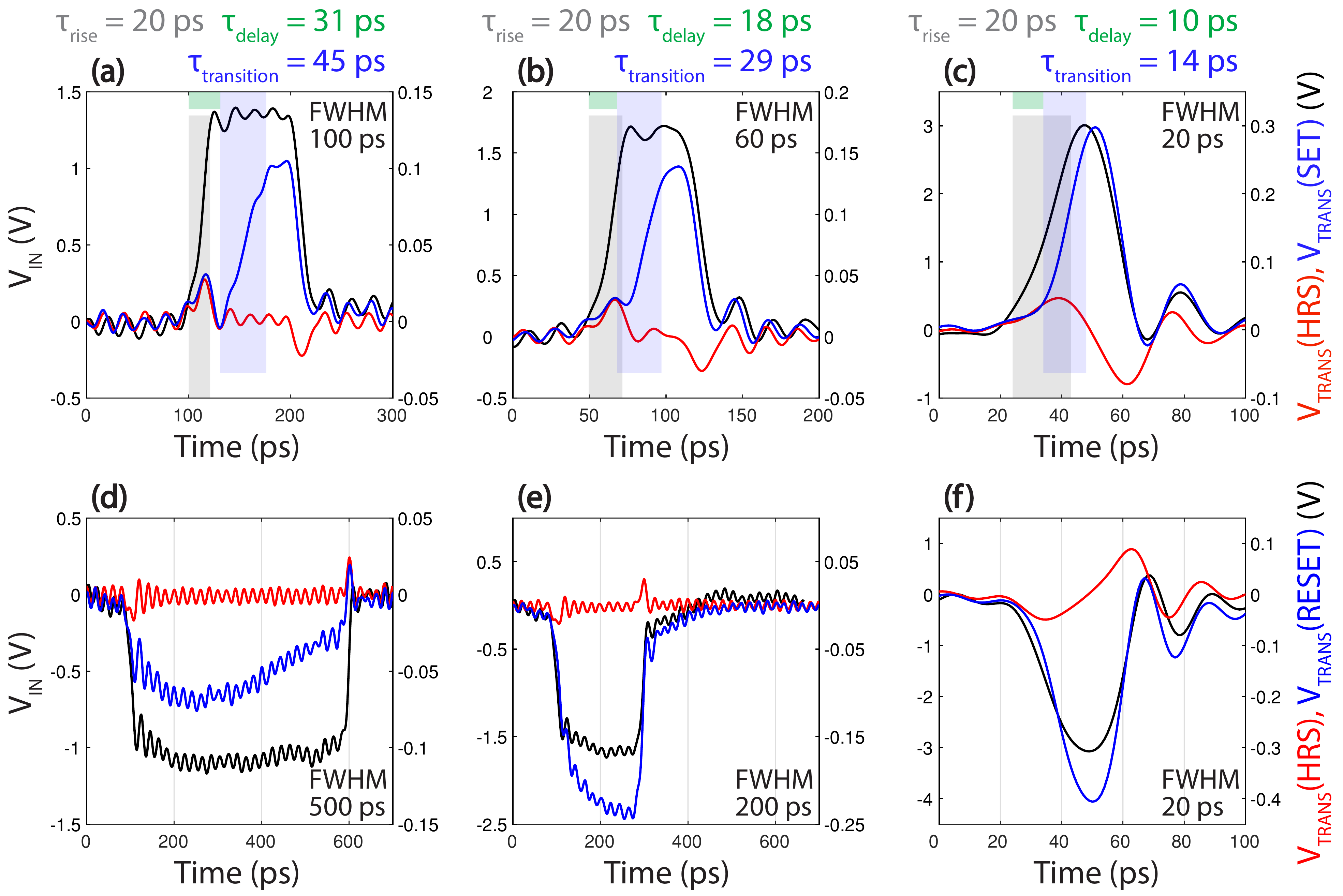}
     \caption{The memristor's resistive response to the fast set and reset pulses are analyzed on the picosecond time-scale. (a-c) Magnified view of the time dependence of $V_{\rm TRANS}$ during the 100~ps, 60~ps and 20~ps long set voltage pulses shown in Fig.~\ref{ns.fig}(a-c), respectively. (d-f) Reset operation due to 500~ps, 200~ps and 20~ps voltage pulses, respectively. For comparison, the measured, predominantly capacitive transmissions of the HRS corresponding to the actual $V_{\rm IN}$ pulses and $C_{\rm MEM}$ = 3~fF are also plotted (red, right axes). The grey shaded areas in (a-c) highlight the $\tau_{\rm rise}$=20~ps rise time of the input pulses defined as the time interval between 10\% and 90\% of the peak value. The blue shaded areas mark the $\tau_{\rm transition}$ time interval of the set switching measured from the apparent divergence of the HRS (red) and switching (blue) traces until 90 \% of the maximum $V_{\rm TRANS}(\rm SET)/V_{\rm IN}$ ratio is reached. The green shaded areas correspond to the $\tau_{\rm delay}$ delay time between the pulse beginning and the onset of resistive switching. The deduced values of $\tau_{\rm rise}$, $\tau_{\rm delay}$ and $\tau_{\rm transition}$ are labeled at the top of the set panels.}
     \label{ps.fig}
\end{figure*}

Finally, the $V_{\rm TRANS}(\rm SET)$ set switching response (blue, right axis) is compared to the above discussed HRS and LRS reference traces in Fig.~\ref{overlay.fig}(c). A clear transition is observed from the non-switching HRS to the LRS traces within a period which is significantly shorter than the $V_{\rm IN}$ pulse duration. Note, however, that the low $C_{\rm MEM}$ is also crucial in terms of identifying the starting point of the transition: should $C_{\rm MEM}$ moderately exceed 3~fF (our devices), the enhanced first peak of the HRS response would `overshadow' the resistive transition until a relatively low $R_{\rm MEM}$ is reached, \textit{i. e.} the divergence of the $V_{\rm TRANS}(\rm HRS)$ and $V_{\rm TRANS}(\rm SET)$ traces, marked by the blue arrow in Fig.~\ref{overlay.fig}(c), would move to the right. Thereby only the last portion of the resistive transition would remain resolvable. Ultimately, when the influence of a large $C_{\rm MEM}$ becomes dominant in $V_{\rm TRANS}(\rm SET)$, the latter can no longer provide any real-time experimental information on the resistive transition. In the present case, at $C_{\rm MEM}$=3~fF the first peak of $V_{\rm TRANS}(\rm HRS)$ restricts the resolvable resistance change window to $\lesssim$20~k$\Omega$ at the shortest available, 20~ps FWHM $V_{\rm IN}$ pulses. Quantitatively corresponding to the noise limitation of our voltage transmission measurement method discussed in Section~2.2, this restriction does not impose any further experimental constraint to our analysis.

\vspace{5mm}
\textbf{2.5 Resistive Switching Time-Scales}
\vspace{5mm}

Next, we apply the above described approach to resolve the corresponding resistance changes during the programming pulses shown in Figs.~\ref{ns.fig}(a-f). The magnified views of the corresponding panels are displayed in Figs.~\ref{ps.fig}(a-f), respectively.

The qualitative behavior of the switching trace (blue, right axes) follows a common pattern at each set pulse duration in Figs.~\ref{ps.fig}(a-c): at the onset of the set pulse (black, left axes) it coincides with the predominantly capacitive HRS response (red trace, right axes) whereas after a transition period it becomes proportional to $V_{\rm IN}$, as expected for a resistive impedance characteristic to the LRS, in accordance with Eq.~\ref{trans.eq}. The time dependence of $V_{\rm TRANS}$ is quantitatively analyzed in terms of the three commonly considered time scales, $\tau_{\rm rise}$, $\tau_{\rm delay}$ and $\tau_{\rm transition}$. The first is an instrumental parameter which is defined between the 10\,--\,90\% of the set pulse peak and equals to $\tau_{\rm rise}$=20$\pm$1~ps, independently of the pulse duration and amplitude, as represented by the grey shaded areas in Figs.~\ref{ps.fig}(a-c). The blue shaded areas correspond to the $\tau_{\rm transition}$ set transition time. This is defined as the time interval from the apparent divergence of the HRS (red) and switching (blue) traces until 90\% of the maximum (resistive) $V_{\rm TRANS}(\rm SET)/V_{\rm IN}$ ratio is reached. Finally, the $\tau_{\rm delay}$ delay time is defined as the time interval between the application of the $V_{\rm IN}$ set pulse and the onset of the $V_{\rm TRANS}(\rm SET)$ set switching response, as illustrated by the green boxes in Figs.~\ref{ps.fig}(a-c). Generally, $\tau_{\rm delay}$ accounts for the generation and initial migration of the ionic species contributing to the filament formation \cite{Yang2014,Witzleben2021b,Siegel2021}. Due to the finite $\tau_{\rm rise}$, the above definition of $\tau_{\rm delay}$ yields to a conservative estimate of the delay time.

The set pulse durations and amplitudes in Figs.~\ref{ps.fig}(a-c) were selected to represent three qualitatively different scenarios. At the application of a 1.3~V amplitude, 100~ps FWHM set pulse [Fig.~\ref{ps.fig}(a)] the $\tau_{\rm transition}$=45~ps long resistive switching starts at $\tau_{\rm delay}$=31~ps, clearly after the set pulse has reached its plateau. This is the regime where the classical voltage-time dilemma, \textit{i. e.}, the exponential acceleration of resistive switching due to a linearly increasing voltage bias applies to the delay time, within the uncertainty imposed by the finite $\tau_{\rm rise}$. At a decreased, 60~ps FWHM but increased, $V_{\rm IN}$=1.7~V amplitude set pulse [Fig.~\ref{ps.fig}(b)] $\tau_{\rm delay}$ is shortened to 18~ps. Thereby the regimes of $\tau_{\rm rise}$ (grey shaded area) and $\tau_{\rm transition}$ (blue shaded area) start to merge, leaving the validity domain of the separately resolvable $\tau_{\rm delay}$ and $\tau_{\rm transition}$ time-constants' description behind. Resistive switching again starts around $V_{\rm IN}$=1.3~V, however the still rising set voltage and the higher, $V_{\rm IN}$=1.7~V plateau result in a faster transition with $\tau_{\rm transition}$=29~ps. Ultimately, at a $V_{\rm IN}$=3~V, 20~ps FWHM pulse [Fig.~\ref{ps.fig}(c)] the three relevant timescales of $\tau_{\rm rise}$, $\tau_{\rm delay}$ and $\tau_{\rm transition}$ largely overlap, facilitating a constantly accelerating set switching between 1.3~V$<V_{\rm IN}<$3~V. The per-definition deduced delay and transition times reach $\tau_{\rm delay}$=10~ps and $\tau_{\rm transition}$=14~ps, respectively.

Below 60~ps FWHM and above $V_{\rm IN}$=1.7~V set voltage pulse amplitude $\tau_{\rm delay}$ falls below $\tau_{\rm rise}$, compromising the adopted, common technical definition of $\tau_{\rm delay}$. However, the observation that the resistive transition is consistently triggered at $V_{\rm IN}\approx$1.3~V suggests that a pure switching threshold voltage description may have a higher practical relevance in this regime. The $\tau_{\rm transition}$=14~ps transition time in Fig.~\ref{ps.fig}(c) corresponds to the $\approx$65~GHz analog bandwidth bottleneck of our high-speed setup. The latter result is a strong experimental evidence that the intrinsic speed limitations of the set switching in valence change memory (VCM) devices must be further pursued in the single-digit picosecond time domain \cite{Bottger2020} or even beyond, corresponding to the phonon frequency \cite{Emboras2018,Menzel2019} of the hosting oxide.

In case of the reset transitions, a closer look at the time dependence of $V_{\rm TRANS}$ in Figs.~\ref{ps.fig}(d-f) reveals a qualitatively different behavior compared to the one of the set switchings. For reference, the non-switching HRS response was also recorded at each programming pulse configuration by adjusting the amplitude of the subsequent set voltage pulses shown in Fig.~\ref{setup.fig}(j) to zero.

The $V_{\rm TRANS}(\rm RESET)$ (blue, right axis) response in Fig.~\ref{ps.fig}(d) is proportional to the $V_{\rm IN}$ (black, left axis) reset pulse of 500~ps FWHM and 1~V amplitude in the first $\approx$80~ps of the pulse duration, signaling a persisting LRS of resistive character. This is then followed by a slow, $\approx$320~ps long period of a gradual transition from a resistive to a capacitive response, as seen by the decreasing magnitude and shifting phase of $V_{\rm TRANS}(\rm RESET)$. The phase can be conveniently verified due to the presence of the small amplitude, 50~GHz modulation of $V_{\rm IN}$. The good quantitative match between the $V_{\rm TRANS}(\rm HRS)$ (red) and $V_{\rm TRANS}(\rm RESET)$ (blue) peaks at 600~ps confirms that the reset transition has been completed within the $V_{\rm IN}$ pulse duration. This behavior, already reported for longer reset pulses \cite{Witzleben2021a}, is qualitatively understood in terms of the simultaneous effects of the electric field and the enhanced Joule heating in the LRS: while the former acts to disassemble the conducting filament, the latter assists this destruction by increasing the ionic mobility of the constituting oxygen vacancies \cite{Marchewka2016}. The reset process is then gradually self-terminated by the reduction of the thermal load due to the increasing device resistance.

In contrast, $V_{\rm TRANS}(\rm RESET)$ shows an entirely resistive response without any trace of a capacitive transition during the 200~ps and 20~ps long reset pulses displayed in Figs.~\ref{ps.fig}(e) and (f). As shown in Figs.~\ref{ns.fig}(e) and (f), the read pulses of either polarity, applied 10~ns before and after the reset pulse unambiguously and reproducibly confirm the completion of the reset switching also for these two reset pulse configurations. Therefore the above observations imply that the actual reset transition takes place in a delayed manner, after the short reset pulse is over. In order to test this hypothesis, we added a probe pulse of positive polarity, 20~ps FWHM and 1~V amplitude to the applied $V_{\rm IN}$ scheme, following the falling edge of the reset pulse at different delay times.

\begin{figure}[t!]
     \includegraphics[width=\columnwidth]{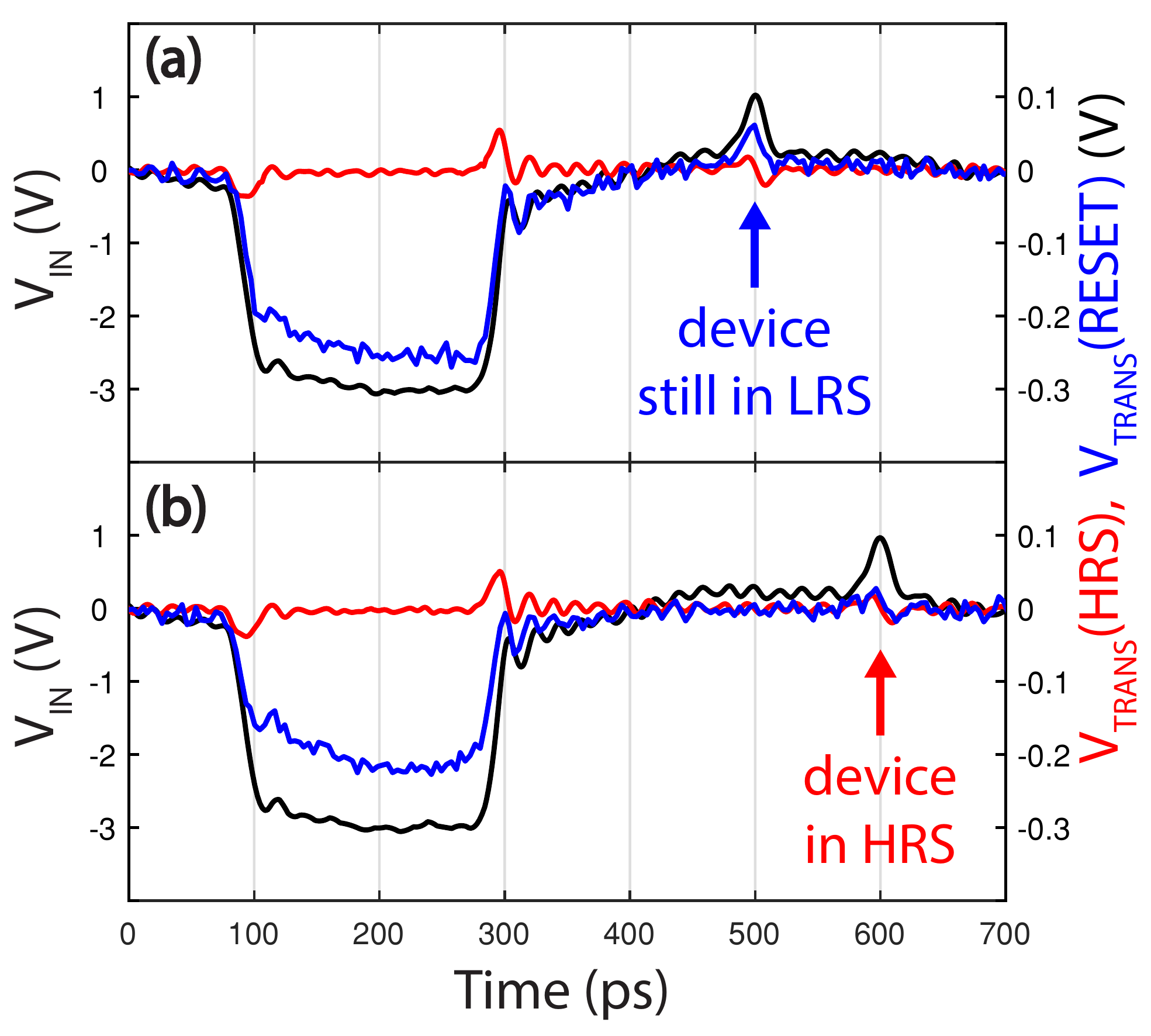}
     \caption{Demonstration of the reset switching delay effect. The $V_{\rm TRANS}(\rm RESET)$ transmission signal (blue, right axes) due to a 200~ps long reset pulse and a subsequent, 20~ps long probe pulse (black, left axes) is shown for different pump-probe pulse delays. The probe pulses, indicated by the arrows, follow the falling edges of the identical reset pulses by 200~ps and 300~ps in (a) and (b), respectively. The device was prepared in identical LRS of $R_{\rm LRS}$ = 2~k$\Omega$. For comparison, the simulated capacitive response corresponding to a HRS with $R_{\rm MEM}$ = 100~k$\Omega$ and $C_{\rm MEM}$ = 3~fF is also plotted (red, right axes).}
     \label{pump-probe.fig}
\end{figure}

\vspace{5mm}
\textbf{2.6 Thermal Reset Delay}
\vspace{5mm}

A representative reset pump-probe experiment is shown in Figs.~\ref{pump-probe.fig}(a) and (b), where the probe pulse was fired 200~ps and 300~ps after the 200~ps FWHM reset pulse, respectively. The device was initialized in similar, $R_{\rm LRS}$ = 2~k$\Omega$ and switched to $R_{\rm HRS} >$ 20~k$\Omega$ according to the 1~ns long read pulses (not shown). The $V_{\rm TRANS}$ response during the reset pulse is entirely resistive in agreement with the previous experiment shown in Fig.~\ref{ps.fig}(e). The $V_{\rm TRANS}$ response to the probe pulse, however, is markedly different depending on the timing of the latter. The probe pulse following the reset pulse by a 200~ps delay triggers a purely resistive response in Fig.~\ref{pump-probe.fig}(a), evidencing that the device is still in its LRS. Delaying the probe pulse another 100~ps longer in Fig.~\ref{pump-probe.fig}(b), on the other hand, results in a capacitive response, in excellent quantitative agreement with the (this time simulated) non-switching HRS transmission corresponding to $R_{\rm MEM}$ = 100~k$\Omega$ and $C_{\rm MEM}$ = 3~fF.

These findings demonstrate that the actual reset transition, although can be triggered with programming voltage pulses even as short as 20~ps, needs longer times, in the order of 100 picoseconds, to complete. Importantly, the discovery of the above described reset delay phenomenon also sheds light on the relevant switching mechanism via the temporal separation of electric field and thermally driven effects. The observation that at the time of the reset transition the electric field has already been turned off for 100's of picoseconds provides a solid evidence that the reset process is predominantly driven by the thermally assisted diffusion of the oxygen vacancies constituting to the conducting filament. Meanwhile, the main role of the reset voltage pulse is to establish the high local temperature via injecting the excessive Joule heat in the LRS. The applied electric field is limited by the instrumental bandwidth and pulse power as well as by the break-down durability of the devices. We propose that the time scale of the thermal delay, on the other hand, can be engineered by the careful thermal design of the samples. The latter is a complex task which must take into account the mesoscopic nature of the heat transport at atomically narrow conducting filaments \cite{Gubicza2015b,Ducry2017} as well as the heat capacitance and thermal conduction of the bulk electrodes and their environment. Device miniaturization is expected to facilitate higher local temperatures and, thus, shorter reset time scales at lower reset pulse power. However, fast cyclic operation would also require fast temperature relaxation which is better served by bulkier metalic pathways for more efficient heat removal. The optimization of the above technological aspects must be addressed by further numerical studies.

We note that time resolved temperature measurements \cite{Stellari2016,Prasad2017} as well as continuum models of heat dissipation and conduction \cite{Reifenberg2006,Reifenberg2008,Xu2013} performed on various nanodevices confirm excess temperatures up to several 100~K as well as decreasing thermal time constants with shrinking active volumes of heat dissipation. Assuming that the heat is released at the scale of the phonon scattering length of $\approx$10~nm \cite{Gubicza2015b} in the metalic LRS of our devices, the reported tendency is consistent with the observed order of 100~ps time-scale.

\vspace{5mm}
\textbf{3. Conclusions}
\vspace{5mm}

By utilizing current state of the art electronics, we presented resistive switching and multilevel programming due to 20~ps FWHM set and reset pulses, confirmed by the evaluation of the steady states before and after the programming pulses. The voltage-time dilemma was also revealed in the sub-100~ps time domain. More importantly, our further analysis was focused on the time dependence of the transmitted voltage at the time-scale of the programming pulses. By the optimization of the device layout, we have shown that the capacitive contribution to the transmission can be sufficiently suppressed even at the fastest bias change rates. This enabled the direct, quantitative comparison of the time dependent transmissions corresponding to the resistive switching transition and to the non-switching HRS and LRS reference states at picosecond time resolution. Thereby we have experimentally demonstrated set delay times below 10~ps and set transition times down to the 14~ps limit of the analog bandwidth of our setup. Our approach allowed us to work with fast and uninterrupted, alternating polarity voltage pulse trains down to 20~ps FWHM in both polarities and to achieve high cycle-to-cycle reproducibility of Ta$_{2}$O$_{5}$ based filamentary resistive switches. 

We have shown that reset switchings can also be reliably triggered by 20~ps FWHM voltage pulses in Ta/Ta$_{2}$O$_{5}$/Pt memristors. However, our pump-probe experiments and the related time dependent analysis also revealed that the duration of the applied reset voltage pulse must not be confused with the actual reset switching time, as the reset process can take an order of magnitude longer time after the ultra-short electrical excitation has been terminated. Such a thermal delay effect, first reported here, provides a strong experimental evidence that the leading mechanism of VCM reset switching is the Joule heating induced thermal diffusion of the oxygen vacancies which can carry on the reset transition also in the absence of an electric field.

Our results underline the importance of the dedicated thermal optimization of the electrode arrangements terminating atomic scale resistive switches for applications requiring their fast cyclic operation. If the reset times are thereby successfully matched with the picosecond-scale set times demonstrated in this work, VCM based filamentary switches can be readily integrated as high-bandwidth, low-power components into state of the art telecommunication technologies and beyond.


\vspace{5mm}
\textbf{4. Methods}
\vspace{5mm}

\begin{figure}[t!]
     \includegraphics[width=\columnwidth]{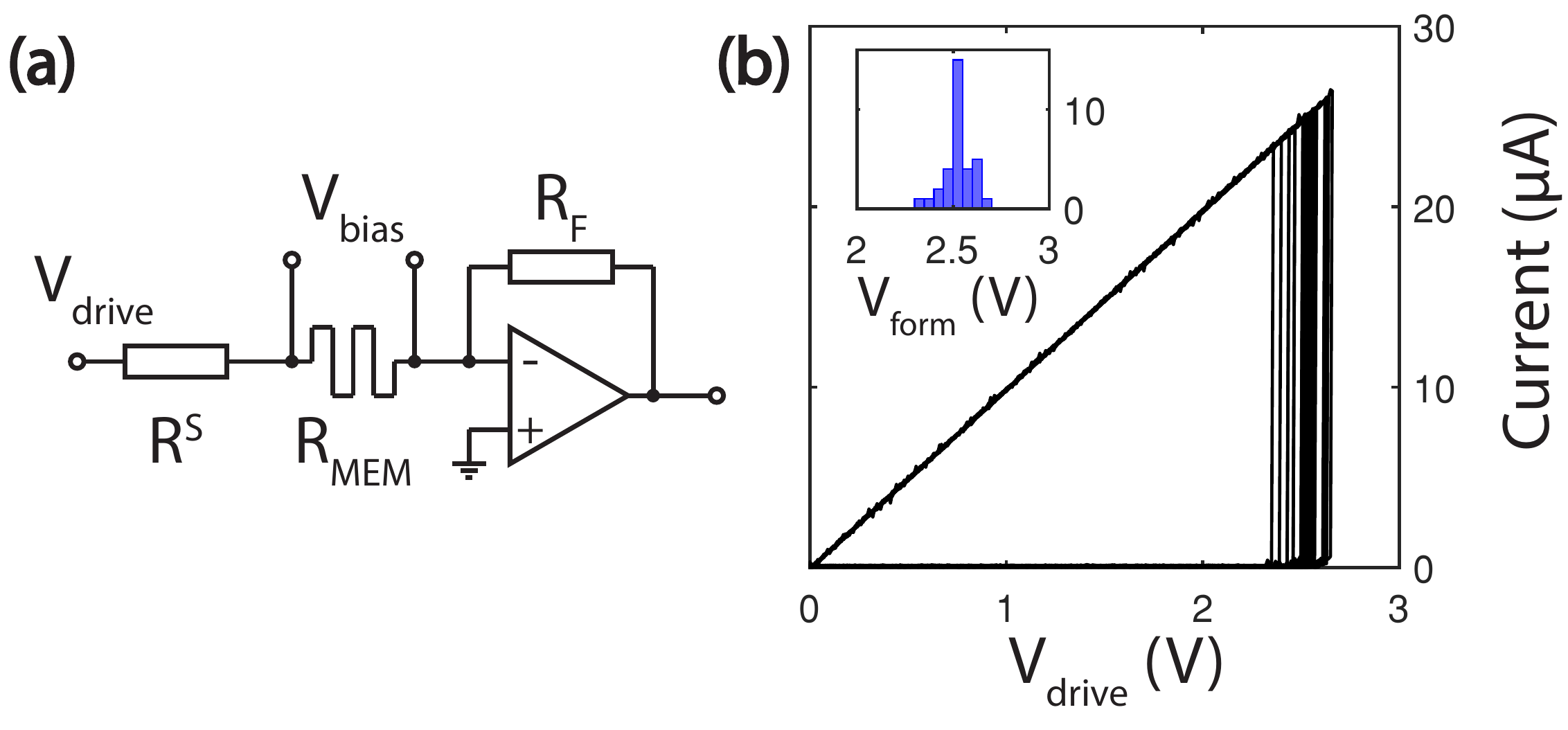}
     \caption{DC characterization of the Ta/Ta$_{2}$O$_{5}$/Pt devices. (a) Schematics of the $I(V)$ measurement setup consisting of a series resistor $R^{\rm S}$, the memristor device and a current amplifier. (b) Electroforming traces of 33 devices recorded with $R^{\rm S}$ = 100~k$\Omega$ limiting the device current. The inset shows the distribution of the forming voltage.}
     \label{dc.fig}
\end{figure}

\begin{figure}[t!]
     \includegraphics[width=\columnwidth]{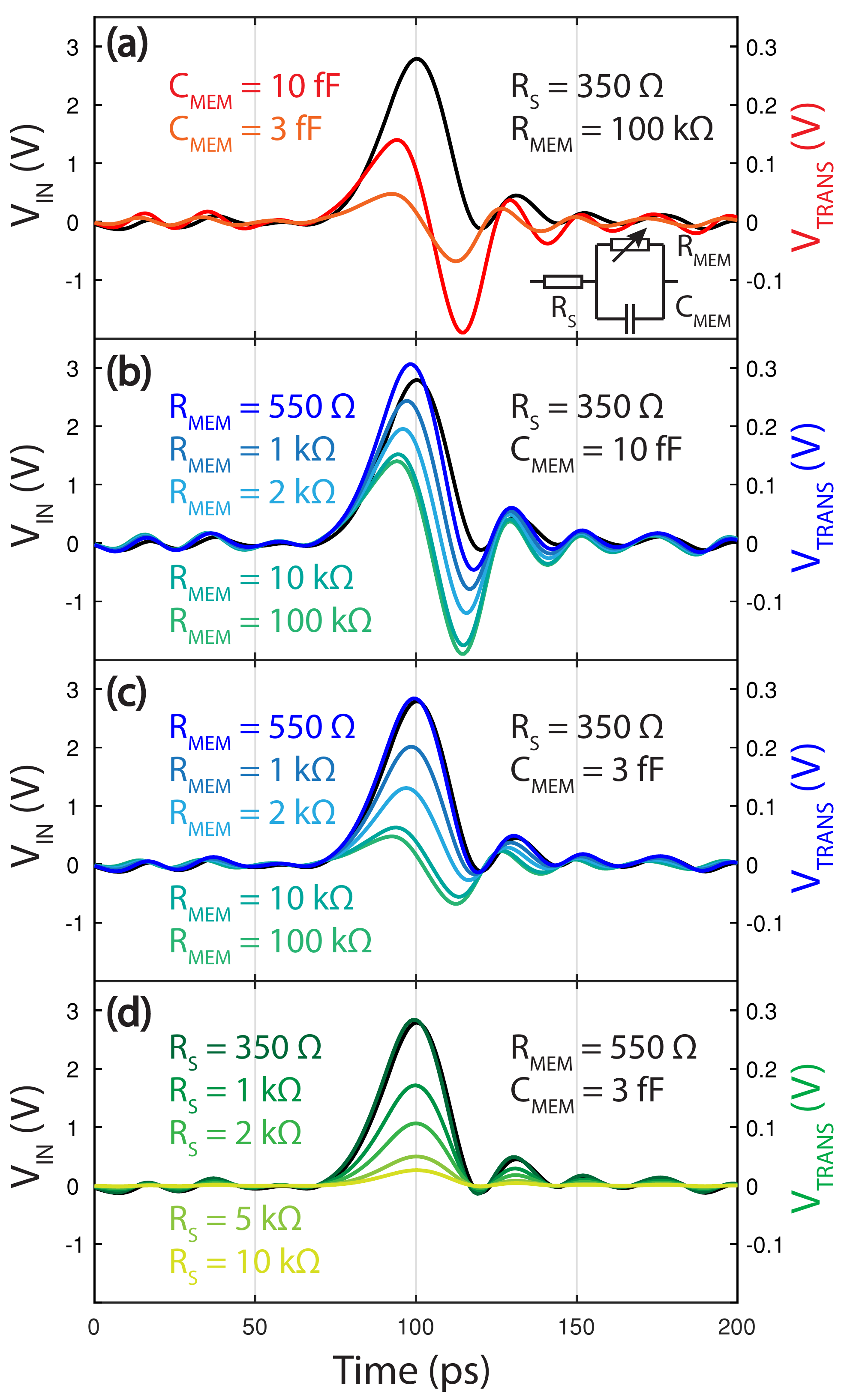}
     \caption{Simulated $V_{\rm TRANS}$ traces for a measured $V_{\rm IN}$ voltage pulse of 20~ps FWHM at various values of the model circuit parameters shown in Fig.~\ref{setup.fig}(h) and also in the inset of (a). The running parameter values indicated to the left of the pulses are displayed in the order of the corresponding traces from top to bottom in each panel. The fixed parameter values are shown on the right-hand side. (a) The effect of the device's parallel capacitance on the HRS response. (b)-(c) The effect of the $R_{\rm MEM}$ memristor resistance at $C_{\rm MEM}$ = 10~fF and $C_{\rm MEM}$ = 3~fF, respectively. (d) The effect of the $R_{\rm S}$ series resistance on the LRS response at $C_{\rm MEM}$ = 3~fF.}
     \label{simulation.fig}
\end{figure}

{\it Sample Fabrication:} The Pt bottom electrodes, the Au terminals of the co-planar waveguide structure and their Ti adhesive layers were deposited by standard electron beam evaporation at a base pressure of 10$^{-7}$~mbar at a rate of 0.1~nm/s. The 5~nm thick Ta$_{2}$O$_{5}$ layers were deposited by reactive high-power impulse magnetron sputtering (HiPIMS) using a Ta target at 6~mTorr pressure, 45~sccm Ar and 5~sccm O$_{2}$ flow rates and 250~W RF power. The stoichiometric composition and the layer thickness were confirmed by XPS spectroscopy. In order to prevent the formation of an ill-defined native oxide at the Ta$_{2}$O$_{5}$/Ta interface, the Ta electrode and its Pt cap were patterned and sputtered on top of the Ta$_{2}$O$_{5}$ film at 4~mTorr pressure, 45~sccm Ar flow and 250~W RF power (125~W dc power) for Ta (Pt).

{\it DC Characterization:} The setup of the dc $I(V)$ measurements is shown in Fig.~\ref{dc.fig}(a). The $V_{\rm drive}$ triangular voltage signal was applied on the memristor sample and the series resistor $R^{\rm S}$ by an NI USB-6341 data acquisition unit (DAQ). The current was measured by a Femto DHPCA-100 current amplifier and the analog voltage input of the DAQ. The $V_{\rm bias}$ voltage drop on the memristor was calculated as $V_{\rm bias}=V_{\rm drive}-I\cdot R^{\rm S}$. During electroforming $R^{\rm S}$ = 100~k$\Omega$ and a 100~mV/s voltage sweep rate were applied. The 10~kSa/s data acquisition frequency and the real-time current feedback prevented excess current loads above 20\,--\,30~$\mu$A. The electroforming traces of 33 devices along with the narrow distribution of the deduced $V_{\rm Form}$ forming voltage values are shown in Fig.~\ref{dc.fig}(b) and its inset, respectively, demonstrating high device-to-device uniformity. Following the above electroforming procedure, the hysteretic $I(V)$ traces were recorded by exposing the memristor and an $R^{\rm S}$ = 1~k$\Omega$ series resistor to slow ($f$=1~Hz) triangular $V_{\rm drive}$ voltage signals.

{\it Transmission Simulations and Sample Layout Considerations:} In order to understand the influence of the individual circuit parameters on the $V_{\rm TRANS}$ response during the ultra-fast voltage pulses, transmission simulations were carried out in LTspice using lossless transmission lines, the equivalent circuit shown in Fig.~\ref{setup.fig}(h) and the measured, 20~ps FWHM $V_{\rm IN}$ pulses. Figure~\ref{simulation.fig}(a) shows the influence of $C_{\rm MEM}$ on the HRS response, represented by $R_{\rm MEM}$ = 100~k$\Omega$. The $R_{\rm S}$ = 350~$\Omega$ series resistance value was experimentally determined by measuring reference samples lacking the Ta$_{2}$O$_{5}$ switching oxide layer. The two red traces corresponding to $C_{\rm MEM}$=3~fF and $C_{\rm MEM}$=10~fF parallel capacitances illustrate the rapidly increasing capacitive character of $V_{\rm TRANS}$ at $C_{\rm MEM}$ values expected from devices exhibiting sub-micron scale active areas. The decisive impact of $C_{\rm MEM}$ is further emphasized in Figs.~\ref{simulation.fig}(b) and (c), where the blue traces span over the same $R_{\rm MEM}$=0.55\,--\,100~k$\Omega$ range at $C_{\rm MEM}$=10~fF and $C_{\rm MEM}$=3~fF, respectively. This quantitative comparison demonstrates how the pronounced capacitive contribution to $V_{\rm TRANS}$ can compromise the experimental resolution of $R_{\rm MEM}$ during the pulse, unless $C_{\rm MEM}$ is carefully minimized. Finally, Fig.~\ref{simulation.fig}(d) underlines that the latter optimization procedure cannot be implemented simply by the aggressive downscaling of the electrode widths, as the resulting increase of $R_{\rm S}$ is also detrimental for the real-time resolution of $R_{\rm MEM}$ during the ultra-fast voltage pulses. Based on the above conclusions, we intuitively implemented the layout shown in Figs.~\ref{setup.fig}(f) and (g), where the active area of the memristor device is confined at the sub-200~nm scale but the narrow section of the leads is also minimized to a 1~$\mu$m distance on either side. The latter feature provided the trade-off between an acceptably low series resistance and a moderate in-plane environmental capacitance. Our co-planar waveguide design adopted the structure outlined in \cite{Torrezan2011}.

\vspace{5mm}
\textbf{Acknowledgements}
\vspace{5mm}

This work was supported by the Werner Siemens Stiftung. M.C. acknowledges financial support from the Swiss National Science Foundation under the Spark Project Nr.~196486. A.H. acknowledges the NKFI K128534 grant.

\vspace{5mm}
\textbf{Conflict of Interest}
\vspace{5mm}

The authors declare no conflict of interest.

\vspace{5mm}
\textbf{Keywords}
\vspace{5mm}

Picosecond resistive switching, thermal reset delay, memristor, conducting filament, Ta$_{2}$O$_{5}$

\vspace{5mm}
\textbf{Author Contributions}
\vspace{5mm}

M.C. conceived the idea of the project with inputs from A.H, I.S., J.L. and U.K. The samples were fabricated by M.C. with contributions from N.J.O. The high-speed experiments were carried out by M.C. with contributions from Y.H. The data was analyzed by M.C. and was discussed with all authors. All authors contributed to the writing of the manuscript.

\bibliography{References}

\end{document}